**One file to share them all: Using the COMBINE Archive and the OMEX format to share all information about a modeling project**

Frank T Bergmann[1] (fbergman@caltech.edu), Richard Adams[2] (ra22597@gmail.com), Stuart Moodie[3,4] (stuart.moodie@eightpillars.uk.com), Jonathan Cooper[5] (jonathan.cooper@cs.ox.ac.uk), Mihai Glont[3] (mglont@ebi.ac.uk), Martin Golebiewski[6] (martin.golebiewski@h-its.org), Michael Hucka[7] (mhucka@caltech.edu), Camille Laibe[3] (laibe@ebi.ac.uk), Andrew K Miller[8] (ak.miller@auckland.ac.nz), David P Nickerson[8] (david.nickerson@gmail.com), Brett G Olivier[9] (b.g.olivier@vu.nl), Nicolas Rodriguez[10] (nicolas.rodriguez@babraham.ac.uk), Herbert M Sauro[11] (hsauro@u.washington.edu), Martin Scharm[12] (martin.scharm@uni-rostock.de), Stian Soiland-Reyes[13] (soiland-reyes@cs.manchester.ac.uk), Dagmar Waltemath[12] (dagmar.waltemath@uni-rostock.de), Florent Yvon[3] (florent@ebi.ac.uk), Nicolas Le Novère[3,10] (lenov@babraham.ac.uk)

**Affiliations**
[1] Modelling of biological processes, BioQUANT/COS, University of Heidelberg, INF 267, Heidelberg, 69120, Germany
[2] ResearchSpace, 24 Fountainhall Road, Edinburgh EH9 2LW.
[3] European Molecular Biology Laboratory, European Bioinformatics Institute (EMBL-EBI), Wellcome Trust Genome Campus, Hinxton, Cambridge CB10 1SD, United-Kingdom
[4] Current affiliation: Eight Pillars Ltd, 19 Redford Walk, Edinburgh EH13 0AG
[5] Department of Computer Science, University of Oxford, Wolfson Building, Parks Road, Oxford, OX1 3QD, United-Kingdom
[6] HITS gGmbH, Schloss-Wolfsbrunnenweg 35, D-69118 Heidelberg, Germany
[7] Computing and Mathematical sciences, California Institute of Technology, Pasadena, CA 91125, USA
[8] Auckland Bioengineering Institute, University of Auckland, Private Bag 92019, Auckland Mail Centre, Auckland 1142, New Zealand
[9] Systems Bioinformatics, VU University Amsterdam, Amsterdam, 1081 HV, The Netherlands
[10] Babraham Institute, Babraham Research Campus, Cambridge CB22 3AT United Kingdom
[11] Department of Bioengineering, University of Washington, Seattle, WA, 98195
[12] Systems Biology and Bioinformatics, University of Rostock, Ulmenstrasse 69, Rostock, 18057, Germany
[13] School of Computer Science, The University of Manchester, Oxford Road, Manchester M13 9PL, United Kingdom

Corresponding author: lenov@babraham.ac.uk

**ABSTRACT**

Background:
With the ever increasing use of computational models in the biosciences, the need to share models and reproduce the results of published studies efficiently and easily is becoming more important. To this end, various standards have been proposed that can be used to describe models, simulations, data or other essential information in a consistent fashion. These constitute various separate components required to reproduce a given published scientific result.

Results:
We describe the Open Modeling EXchange format (OMEX). Together with the use of other standard formats from the Computational Modeling in Biology Network (COMBINE), OMEX is the basis of the COMBINE Archive, a single file that supports the exchange of all the information necessary for a modeling and simulation experiment in biology. An OMEX file is a ZIP container that includes a manifest file, listing the content of the archive, an optional metadata file adding information about the archive and its content, and the files describing the model. The content of a COMBINE Archive consists of files encoded in COMBINE standards whenever possible, but may include additional files defined by an Internet Media Type. Several tools that support the COMBINE Archive are available, either as independent libraries or embedded in modeling software.

Conclusions:
The COMBINE Archive facilitates the reproduction of modeling and simulation experiments in biology by embedding all the relevant information in one file. Having all the information stored and exchanged at once also helps in building activity logs and audit trails. We anticipate that the COMBINE Archive will become a significant help for modellers, as the domain moves to larger, more complex experiments such as multi-scale models of organs, digital organisms, and bioengineering.

**KEYWORDS**

**BACKGROUND**

The ability to obtain similar results when reproducing an experiment is a tenet of modern science. Reproducible science has even become a field of research in its own right [1,2]. The reproducibility of a scientific study depends on the careful description of the original experiment, including the methods and tools used to perform the experiment, the substrate on which to perform the experiment, and the precise experimental setup, including all necessary influences from the environment. When the result is meant to be presented after post-processing, it is also imperative to provide the details of the processing steps. With the rise of systems biology, computational models have been increasingly used for *in silico* experiments and this has led to a corresponding need to reproduce the results reliably.

The need for exchanging mathematical model structure - that is the list of variables and how they are related - was recognized early on. It led to the generation of tool-independent structured formats, such as SBML [3], CellML [4], and NeuroML [5] to cite only a few. While the exchange of model structure helped with model reuse, the research community recognized that more contextual information was required to fully reproduce the results obtained from a model. Several guidelines were developed to list this information, including the Minimum Information Requested in the Annotation of Models (MIRIAM) [6] and the Minimum Information About Simulation Experiments (MIASE) [7]. Further structured formats were needed to implement these guidelines, such as SBRML [8] for result data and SED-ML [9] for the description of simulation setups.

The modeling and simulation life cycle includes several steps [10] which need to be documented and described in order to be reproduced. Typically, a model is built based upon prior knowledge, either manually or through automated means. Although not generally encoded in a standard form, the description of the procedure used to build it, including accompanying hypotheses, assumptions and approximations, is often crucial to guarantee proper use of a model [11]. The model itself (the variables, their interrelationships and the constraints) must be precisely described. In addition, a precise description of simulation and analysis procedures must also be given in order to allow them to be reproduced. Simulation results might be needed as well, either for comparison to other data or as models for further analysis (for instance individuals generated from a population model). One may also want to add other information, for instance experimental result data (to fit the model or representative simulation results) encoded in NuML [12] or FieldML [13], model structure represented in the SBGN [14] and encoded in SBGN-ML [15], or a description of the biological system being modeled, such as a synthetic biology construct encoded in SBOL [16]. Therefore in order to fully reproduce a modeling and simulation procedure, one needs to access a constellation of files in different formats, including model and simulation descriptions, numerical data  and links to additional information (metadata).

The increasing complexity of systems under study has led to the development of modular approaches, where models are built from several files describing different parts of the model. This is, for instance, the case for models encoded using CellML, NeuroML, or the SBML Level 3 Hierarchical Model Composition package [17]. Furthermore, other fields in the life sciences, such as ecology [18] and drug development [19], have also started to develop standards to encode their specialized mathematical models. In these fields, datasets and simulation procedures are integral parts of the model descriptions. All the files necessary for the description of a model must therefore be exchanged in order to fully understand and use the model.

Managing multiple files presents difficulties. Sharing those files can be a tedious and error-prone affair. Some files necessary to build or process a model might have moved, or even be deleted, precluding the reproduction of results and ultimately even re-using the model. Another difficulty with handling multiple files is that models evolve over time. Some changes, such as corrections of parameter values or changes in the network structure, affect the outcome of the scientific study. Corrections of previously published model code must be communicated and propagated to all instances of the model that have been reused in other studies. It is therefore necessary to specify the version of a model that has been used in a simulation [20]. Solutions towards improved management of computational models have recently been proposed, including better accessibility of models and versions thereof, and better links between all files related to a specific study [20-24]. However, all these approaches focus on long-term availability and management of modeling experiments. They do not provide the means for easy export and exchange of models in structured and well-defined formats. One advantage of an archive is that a single file can encapsulate everything there is to know about a specific modeling project, including the instructions on how to "open" the archive and interpret it. Similar examples from the domain of computer science are the packages of the Debian GNU/Linux operating system [25], the Java Archives [26], the Microsoft Office Open XML [27], and the Open Document Formats [28].

In this article, we describe a type of archive developed by the COMBINE initiative [29] that enables the exchange of all information required to reproduce a modeling project.

## IMPLEMENTATION

### Format of the archive

The COMBINE Archive is encoded using the "Open Modeling EXchange format" (OMEX). The archive itself is a "ZIP" file [30]. ZIP is used for data compression and archiving purposes. A ZIP file contains one or more files that have been compressed, to reduce file size, or stored as is (Figure 1). The technical specification of the ZIP format is available from the PKWARE website [31]. The default file extension for the COMBINE Archive is *.omex*. Additional extensions are available to indicate what is the main standard format used within the archive. This helps users choose between different archives, and select appropriate software tools with which to open them.

- *.sedx* - SED-ML archive
- *.sbex* - SBML archive
- *.cmex* - CellML archive
- *.sbox* - SBOL archive
- *.neux* - NeuroML archive
- *.phex* – DDMoRe archive using the PharmML format

Note that a COMBINE Archive may contain files in several different standard formats. Therefore the specific file extension is only an indication provided for the convenience of the user. For instance, a database of models could distribute archives of the same models encoded in SBML *.sbex* and CellML *.cmex*. However, archives containing models in both formats could also be distributed and the extension *.omex* used. If the archives contain SED-ML files, the extension *.sedx* could be used and the selection between the model formats be devolved to the SED-ML file.

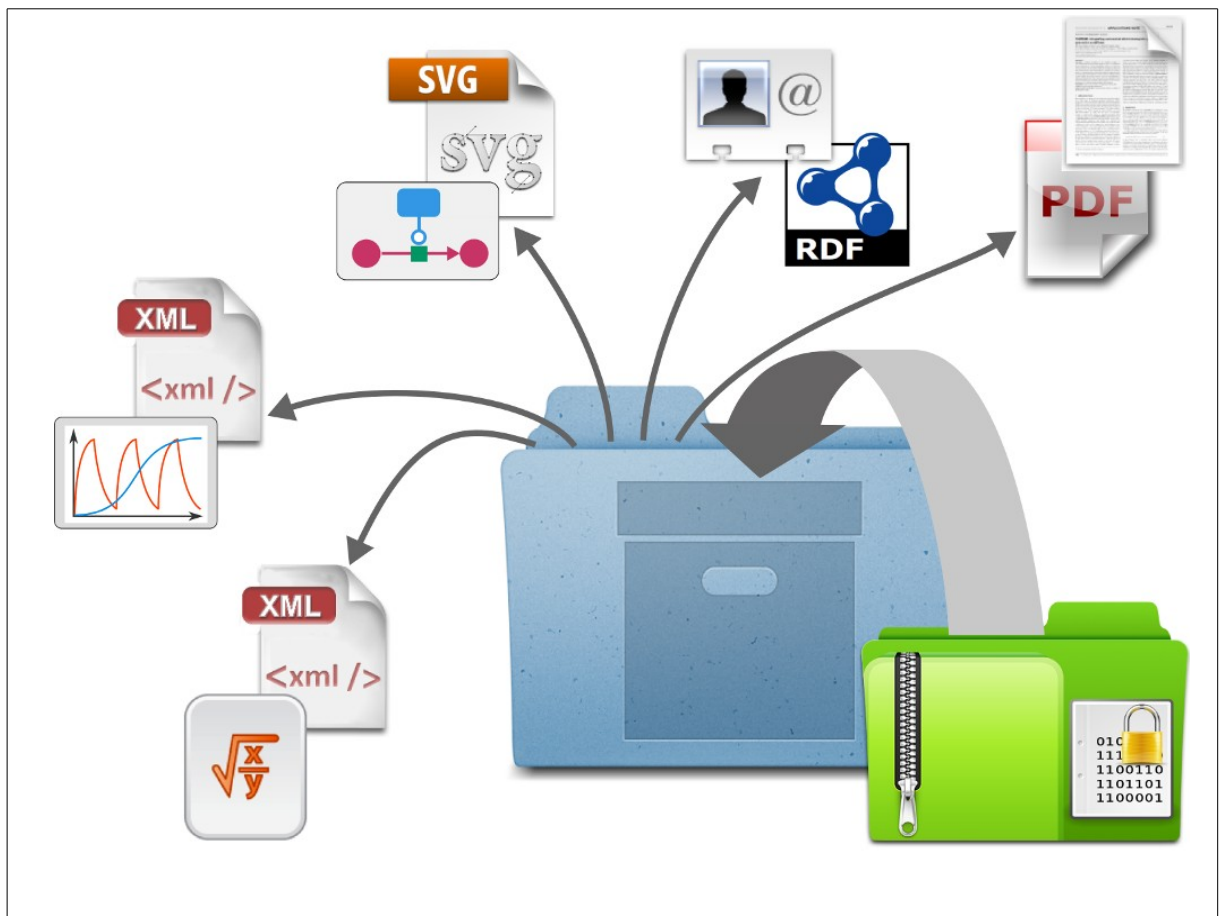

**Figure 1**: Schematic representation of the structure of a COMBINE Archive, a (possibly encrypted) zipped version of an archive containing descriptions of models, simulations, graphical representations, metadata and other sources of information.

**Manifest file**

Every COMBINE Archive contains at least one file, located at the root, i.e. highest in the hierarchy of files inside the archive. This mandatory file is called *manifest.xml*. It is an XML file that contains a flat list of names of all files constituting the archive, and it describes each file's type and location inside the archive. Figure 2 shows an example manifest file. A valid manifest file must have at least one entry, the one for the archive itself (first content entry in Figure 2). However, the manifest may contain as many entries as needed. In the current version of OMEX, all the files listed in the manifest must be included in the archive itself. The location of those files is defined by a relative Uniform Resource Identifier (URI) [32].

Each entry is encoded in an XML element named <content>, and characterized by three attributes. The attribute *location* contains a URI that specifies where the file is located with respect to the root of the archive. The attribute *format* specifies the file format, using either an Identifiers.org URI [33] if it exists (i.e. for any format part of COMBINE, e.g. *http://identifiers.org/combine.specifications/sed-ml.level-1.version-2*), or an Internet Media Type URI [34] (e.g. *http://purl.org/NET/mediatypes.application/pdf*). If a format is neither registered with COMBINE nor with IANA, an unregistered Internet media type can be used. Such unregistered Media Type should take the form *type/x.name* (i.e. using URIs such as

*http://purl.org/NET/mediatypes.application/x.matlab* or *http://purl.org/NET/mediatypes.application/x.copasi*). Finally, the optional *master* attribute is a Boolean which indicates that the content should be considered the main one, to be read or presented to a user. The *master* attribute, read by the processing software once the archive is loaded and parsed, is different from the extensions described above. For instance, the manifest of an archive containing a simulation description and all the models necessary to run a simulation, could have the *master* flag set to "true" for the SED-ML file. In an archive containing a modular model made up of many parts that are hierarchically linked, the *master* flag should be set to "true" for the master model, which in turn is importing sub-models but is not called by any.

```xml
<?xml version="1.0" encoding="utf-8"?>
<omexManifest xmlns="http://identifiers.org/combine.specifications/omex-manifest">
    <content location="."
        format="http://identifiers.org/combine.specifications/omex"/>
    <content location="models/model.xml"
        format="http://identifiers.org/combine.specifications/sbml"/>
    <content location="simulation.xml" master="true"
        format="http://identifiers.org/combine.specifications/sed-ml"/>
    <content location="doc/article.pdf"
        format="http://purl.org/NET/mediatypes/application/pdf"/>
    <content location="metadata.rdf"
        format="http://identifiers.org/combine.specifications/omex-metadata"/>
</omexManifest>
```

**Figure 2:** Example of manifest file describing 5 files: the manifest itself, an SBML file with the structure of the model, a SED-ML file with the description of a simulation, a PDF version of the article describing the modeling and simulation experience, and an RDF file containing metadata about the archive. In this specific archive, the SED-ML file should be read first, which is indicated by the master attribute set to "true" on the file *simulation.xml*.

**Metadata**

Any type of file can be included in a COMBINE Archive, and therefore any type of metadata format may be used to encode clerical information or add semantics to the modeling contents in the archive. However, in the interests of interoperability and to ease the development of software support for metadata, the archive specification recommends an XML serialization of the Resource Description Framework [35] contained in a file called *metadata.rdf* (Figure 3). The format reuses several existing standard terminologies:

- The Resource Description Format itself.
- vCard 4 [36], a standard for electronic business cards; in particular its terms *hasName, given-name, family-name, hasEmail, organization-name.* How to use vCard in RDF is specified by the vCard ontology, a recommendation from the W3C [37].
- Metadata terms of the Dublin Core Metadata Initiative [38], in particular the terms *description, creator, created, modified, W3CDTF* (to encode a date, see [39]). More information on the use of Dublin Core in RDF can be found on the Dublin Core website [40].

```xml
<?xml version="1.0" encoding="UTF-8"?>
<rdf:RDF xmlns:rdf="http://www.w3.org/1999/02/22-rdf-syntax-ns#"
         xmlns:dcterms="http://purl.org/dc/terms/"
         xmlns:vCard="http://www.w3.org/2006/vcard/ns#"
         xmlns:bqmodel="http://biomodels.net/model-qualifiers/">
    <rdf:Description rdf:about=".">
        <dcterms:description>
            Expanded version of the human metabolic reconstruction Recon 2.1
        </dcterms:description>
        <dcterms:creator rdf:parseType="Resource">
            <vCard:hasName rdf:parseType="Resource">
                <vCard:family-name>Le Novere</vCard:family-name>
                <vCard:given-name>Nicolas</vCard:given-name>
            </vCard:hasName>
            <vCard:hasEmail rdf:resource="lenov@babraham.ac.uk" />
            <vCard:organization-name>
                Babraham Institute
            </vCard:organization-name>
            <vCard:hasURL rdf:resource="http://orcid.org/0000-0002-6309-7327"/>
        </dcterms:creator>
        <dcterms:created rdf:parseType="Resource">
            <dcterms:W3CDTF>2014-06-26T10:29:00Z</dcterms:W3CDTF>
        </dcterms:created>
        <bqmodel:is
            rdf:resource="http://identifiers.org/biomodels.db/MODEL1311110001" />
        <bqmodel:isDescribedBy
            rdf:resource="http://identifiers.org/arxiv/1311.5696" />
    </rdf:Description>
</rdf:RDF>
```

**Figure 3:** Example of a metadata file bringing additional information about the archive itself (identified by the relative path "."): a short description of what the archive is about, details about its creator, and dates of creation and last modification.

The metadata file should provide sufficient information to follow the MIRIAM and MIASE guidelines whenever possible. At the very least, the metadata file should provide the archive creation date, the date of last update, and who created it. COMBINE Archive creators could also provide a description of the software tool that generated the archive, as well as a reference to external information describing the work. In addition to the information about the archive itself, one could provide metadata about any of the archive content files. This information helps building logs of actions performed as well as audit trails, which are, for instance, important for models used in drug development. Future development of the metadata content and format is the topic of discussions of the combine-metadata Google group.

**RESULTS AND DISCUSSION**

The SED-ML community initially developed the concept behind the COMBINE Archive, when faced with the need to encapsulate the simulation experiment description and the models needed to perform that experiment [41]. Members of the community extended the aim of the SED-ML archive to encompass any file type that would be useful during a modeling and simulation procedure, and wrote an initial technical description. This draft was discussed during several COMBINE meetings, on the combine-discuss mailing list and subsequently on the dedicated combine-archive mailing list. The discussions converged towards the specification described in this manuscript. A first release candidate was published on July 4 2014. Comments were taken into account in a second release candidate published on July 25 2014. No further comments on the document were received and OMEX version 1 was released on September 15[th] 2014. Its specification is described in a document available at `http://identifiers.org/combine.specifications/omex`.

Several tools providing support for the COMBINE Archive have already been released (Table 1) either as independent libraries or embedded within modeling software such as PySCeS [42], VCell [43] or Tellurium [44].

| Name | Main developer | Language | URL |
|---|---|---|---|
| LibCombine | Frank Bergmann | C# | https://github.com/fbergmann/CombineArchive |
| CombineArchive Toolkit | Martin Scharm | Java | https://sems.uni-rostock.de/ |
| libCombineArchive | BioModels team | Java | https://github.com/mglont/CombineArchive |
| PySCeS | Brett Olivier | Python | http://pysces.sourceforge.net/ |
| Tellurium | Herbert Sauro | Python | http://tellurium.analogmachine.org |
| VCell | Ion Moraru | Java | http://www.nrcam.uchc.edu/ |

**Table 1:** Different implementations that support the COMBINE Archive today. [Table 1 here]

**Use cases**

Combining several types of information encoded in different formats into a single file will be useful for many researchers. Here we showcase a few applications of the COMBINE Archive. These examples demonstrate the current usage of the archive and are not meant to be limiting.

Use case 1, Systems biology models

As recognized by the MIRIAM and MIASE guidelines, the sole description of model variables and their relationships (the structural model) is not sufficient to allow the reproduction of simulation results. The description of the simulation and analysis tasks is also necessary. Many modeling and simulation software configuration formats include both model description and experiments. In order to exchange this information in standard, tool-independent, formats, one must provide the models (for instance encoded in SBML) and the simulation description (in SED-ML). An example of such archive is provided in Supporting data 1. Although a SED-ML file can describe an experiment using models available remotely (using URIs to allow for their identification and retrieval), it is often useful to be able to provide all the necessary information at once. One might also need several models to reproduce a simulation experiment, for instance to compare results generated from different models, to parametrize a model using the results of another model or to run several models concurrently (see also use case 3). Moreover, a model can be described in multiple documents, for instance when encoded in CellML 1.1 [45] or in SBML Level 3 with the model composition package [17]. The COMBINE Archive permits researchers to share all the documents describing a model and associated simulations in a single file.

Use case 2, Drug discovery models

Assessing the effects of a drug using mathematical models involves several steps, including model selection, parameter estimation, population simulations etc., both on the pharmacokinetics (drug concentration over time) and pharmacodynamics (drug effect versus drug concentration) side. The US Food and Drug Administration requires the provision of information sufficient to completely reproduce the evaluation of a drug [46]. This includes SAS transport files [47] to represent all datasets used for model development and validation; ASCII text files of model codes or control streams and output listings for all major model building steps (base structural model, covariates models, final model, and validation model); individual plots for a representative number of subjects for population analysis; and standard model diagnostic plots. Each data item must be accompanied by a description provided in a PDF file. A structured archive containing all these components is a convenient method of ensuring that all information is faithfully transmitted, and that the correct versions of each piece of information are included in the transmission. PharmML, the emerging standard for pharmacometrics models, already supports the COMBINE Archive as a container that encapsulates all the information pertaining to a pharmacometrics modeling project [19].

Use case 3, Large hybrid modular models

As systems biology moves toward the description of more complex systems, such as comprehensive biological processes [48], whole cells [49], and organs [50], larger and more detailed models are being developed. These models encompass biological processes that might require the use of different modeling approaches. Their simulations sometimes require the use of several simulation tools. The tools' results influence each other, for instance using synchronization [49, 51]. To be able to reproduce such simulation experiments, one must provide all the submodels, all the simulation descriptions, and the descriptions of the overall experiments with the coordination of the elementary simulations. All files must be at the right place and in the right version. A single file archive offers a convenient way to share consistent instances of those models.

Use case 4, Automatic (machine-only) transfer of research results

Given a suitable technical infrastructure, no human interaction is necessary to transfer tasks and results between machines or applications. The COMBINE Archive can be seen as a container for all files necessary for the job to run a simulation on a computational model. The archive can be submitted to compute nodes which can then automatically read and process the corresponding tasks. For example, the Functional Curation project of Chaste [11,52] uses the COMBINE Archive as the standard format to transfer data between the web interface and the backend. The goal of the Functional Curation project is to compare a model's behavior under different experimental scenarios. On the web page, a user may choose the set of computational models of interest together with a set of experiment descriptions they would like to test against these models. All selected files are compiled into a COMBINE Archive, for instance using the CombineArchive library [53], and sent to a node in a back-end, which is able to understand the encoded job and run the experiment. Afterwards, the simulation results are again packed into a COMBINE Archive and sent back to the web server where they are presented to the user. Thus, the COMBINE Archive eases the communication between nodes in a network.

**Related efforts**

The idea of a single package to run a simulation experiment in biology was pioneered by the JSim "Project File", which are text files that encapsulate everything used by the JSim program: the notes, the model code, and the control parameters for all the steps in the analysis [54]. More recently, the Workflow4Ever project (Wf4ever), developed the Research Object Bundle [55], focusing on the preservation of scientific experiments in data-intensive science [56]. The structure of the Research Object Bundle is close to the COMBINE Archive. Based on Adobe *Universal Container Format* (UCF) [57], it is also a ZIP file containing a manifest and metadata. One way to start converging the efforts is to share metadata vocabularies and formats, as demonstrated in the tool ro-combine-archive [58], which enriches a COMBINE Archive such that it can co-exist as a Research Object Bundle and vice versa.

Beyond scientific research, several industry standards are being widely used to build document archives based on ZIP files. They mostly belong to two large families. The aforementioned UCF bears strong similarities with the *Open Document Format* [28] developed by the Organization for the Advancement of Structured Information Standards (OASIS), and the *EPUB Open Container Format* (OCF) [59] developed by the International Digital Publication Forum (IDPF). As with OMEX, those archives contain a *manifest* file that lists and identifies the contents of the archive and a *metadata* file, either in XML or in RDF. In parallel, Microsoft developed the *Open Packaging Convention* [60] used in its Office Suite and some other software such as MathWorks' Simulink. Those packages also list their contents in an XML file, and can carry metadata .

However, these industry standards can be relatively complex and rigid. In this initial iteration of the OMEX specification, the community opted for simplicity and flexibility, in order to encourage rapid support and usage. The main differences between the COMBINE Archive and the formats discussed above lie in the structure of the archive and the format of the file describing the content. Since the COMBINE Archive is in general less restrictive, future convergences will be easy to implement once agreed upon.

**General discussion**

A COMBINE Archive contains files encoded in various formats. Currently OMEX does not provide a standard mean to describe the relationships between the different files beyond the mechanisms used by the various file encoding formats. For example, SED-ML documents define links to the models and datasets it uses; Similarly, hierarchical model documents in CellML or SBML contains links to the required sub-models. Some work remains to be done in order to specify the relationship between given elements of a file with elements in another file. For instance, there is currently no well defined way to link glyphs encoded in SBGN-ML with the corresponding elements of an SBML file. Some solutions have been discussed, including systems similar to the SED-ML link to models or more general mapping files.

In addition to the multi-files packaging aspect, the use of the ZIP format with OMEX allows for significant compression of file contents and resultant reduction in file size. This provides a notable benefit when sharing state-of-the-art models and associated files, which can be quite large. For instance an archive containing the expanded version of the human metabolic reconstruction Recon2.1 [61] is approximately 23MB bytes, while the sum of its contained files is 230MB, i.e. 90% compression ratio (Supporting data 2).

The current specification requires that all the components are encapsulated within the archive. Accessing documents outside the archive might be more flexible and fit with a semantic web approach. However, it would require more complex verification and validation systems to answer questions such as "Are the links still valid? Are the versions of the files meant to be used in the archive the same as the version accessible through the URLs? How do users and software tools discriminate between network failure, non-existent file, file in the wrong format, etc.?". For the time being, it was felt that including everything in the archive file was a better choice. If an archive cannot be read or interpreted, one can attribute the fault to either the tool reading the archive or the archive builder. The reliability of third parties is inconsequential in this scheme.

## CONCLUSIONS

The ability to distribute an entire modeling project as a single file, containing all the necessary protocols and data needed for its implementation, will lead to easier reuse of models, improved reproducibility of results, and better tracking of the model development process.

## AVAILABILITY AND REQUIREMENTS
The specification of the COMBINE Archive is available at:
http://identifiers.org/combine.specifications/omex

## AVAILABILITY OF SUPPORTING DATA
The data set supporting the results of this article is included within the article (and its additional file).

## LIST OF ABBREVIATIONS USED

**COMBINE** Computational Modeling in Biology Network
**MIASE** Minimum Information About a Simulation Experiment
**MIRIAM** Minimum Information Required in the Annotation of Models
**NuML** Numerical Markup Language
**OCF** Open Container Format
**ODF** Open Document Format
**OMEX** Open Modeling EXchange format
**OOXML** Office Open XML
**OPC** Open Packaging Convention

**SBGN** Systems Biology Graphical Notation
**SBGN-ML** Systems Biology Graphical Notation Markup Language
**SBML** Systems Biology Markup Language
**SBRML** Systems Biology Result Markup Language
**SED-ML** Simulation Experiment Description Markup Language
**UCF** Universal Container Format
**URI** Uniform Resource Identifier

## COMPETING INTERESTS

Authors declare no competing interests.

## AUTHORS' CONTRIBUTIONS

FTB, RA, NLN wrote the initial version of COMBINE Archive, JC, MG, MH, AKM, SM, DPN, SSR, NR, HMS, DW contributed to the current specification, SM, MG, BO, MS, FY, SSR developed implementations. All authors contributed to the manuscript.

## ACKNOWLEDGEMENTS

SM, MG, CL received funding from the Innovative Medicines Initiative Joint Undertaking (grant agreement no. 115156, DDMoRe). JC was supported through the EPSRC Cross-Disciplinary Interface Programme (grant number EP/I017909/1) and Microsoft Research Ltd. MH and NR were supported by US National Institute of General Medical Sciences (GM070923). DPN is supported by The Virtual Physiological Rat Project (NIH P50-GM094503) and the Maurice Wilkins Centre for Molecular Biodiversity. BO is supported by a ZonMW Zenith grant 40-41009-98-10038. HMS was partly supported by US National Institute of General Medical Sciences (R01GM081070). MS and DW are funded by the German Federal Ministry of Education and Research (e:Bio program SEMS, FKZ 031 6194). SSR was supported by the EU (FP7 RI-283359 and FP7-ICT-2007-6 270192). NR and NLN were supported by the UK Biotechnology and Biological Science Research Council (BBS/E/B/000C0419).

**ADDITIONAL FILE**

**repressilator.omex** - COMBINE Archive containing a model of the Repressilator encoded in SBML and the simulation experiment description provided as supporting information for [9]. Available at http://lenoverelab.org/sites/default/files/documents/repressilator.omex

**recon2-1-x.omex** - COMBINE Archive containing the expanded version of the Recon1.1 model. Available at http://lenoverelab.org/sites/default/files/documents/recon2-1-x.omex